\documentclass[twocolumn]{aastex62}

\usepackage[all]{hypcap} 
\def\sectionautorefname~#1\null{\S#1\null}
\def\subsectionautorefname~#1\null{\S\S#1\null}
\def\subsubsectionautorefname~#1\null{\S\S\S#1\null}

\usepackage[utf8]{inputenc}
\usepackage[T1]{fontenc}

\newcommand{\todo}{\ifmmode \text{\color{red}\Huge{\(\bullet\)}} \else {\color{red}{\Huge$\bullet$}}\fi}
\newcommand{\et}{et al.\ }
\newcommand{\ld}	{\ifmmode {\rm l.d.} \else l.d.\fi}
\newcommand{\kms}	{\ifmmode {\rm km\,s}^{-1} \else km\,s$^{-1}$\fi}
\newcommand{\cc}	{\ifmmode {\rm cm}^{-3}    \else cm$^{-3}$\fi}
\newcommand{\cmii}	{\ifmmode {\rm cm}^{-2}    \else cm$^{-2}$\fi}
\newcommand{\ergs}	{\ifmmode {\rm erg\,s}^{-1} \else erg s$^{-1}$\fi}
\newcommand{\ergcms}	{\ifmmode {\rm erg\,cm}^{-2}\,{\rm s}^{-1} \else erg\,cm$^{-2}$\,s$^{-1}$\fi}
\newcommand{\ergcmsA}	{\ifmmode {\rm erg\,cm}^{-2}\,{\rm s}^{-1}\,{\rm\AA}^{-1}
\else erg\,cm$^{-2}$\,s$^{-1}$\,\AA$^{-1}$\fi}
\newcommand{\kev}	{\ifmmode {\rm keV} \else keV\fi}
\newcommand{\Msun}{\ifmmode M_{\odot} \else $M_{\odot}$\fi}
\newcommand{\Lsun}{\ifmmode L_{\odot} \else $L_{\odot}$\fi}
\newcommand{\Zsun}{\ifmmode Z_{\odot} \else $Z_{\odot}$\fi}
\newcommand{\mpyr}{\ifmmode \Msun\,{\rm yr}^{-1} \else $\Msun\,{\rm yr}^{-1}$\fi}
\newcommand{\Msol}{\Msun}

\newcommand{  \Lya      }{\ifmmode {\rm Ly}\alpha \else Ly$\alpha$\fi}
\newcommand{  \Halpha   }{\ifmmode {\rm H}\alpha \else H$\alpha$\fi}
\newcommand{  \halpha   }{\Halpha}
\newcommand{  \ha       }{\Halpha}
\newcommand{  \Hbeta    }{\ifmmode {\rm H}\beta \else H$\beta$\fi}
\newcommand{  \hbeta    }{\Hbeta}
\newcommand{  \hb       }{\Hbeta}
\newcommand{  \heii     }{\ifmmode {\rm He}\,\textsc{ii} \else He\,\textsc{ii}\fi}
\newcommand{  \HeIIop   }{\ifmmode {\rm He}\,\textsc{ii}\,\lambda4686 \else He\,\textsc{ii}\,$\lambda4686$\fi}
\newcommand{  \ciii     }{\ifmmode {\rm C}\,\textsc{iii}\right] \else C\,\textsc{iii}]\fi}
\newcommand{  \CIII     }{\ifmmode {\rm C}\,\textsc{iii}\right]\,\lambda1909 \else C\,\textsc{iii}]\,$\lambda1909$\fi}
\newcommand{  \civ      }{\ifmmode {\rm C}\,\textsc{iv}  \else C\,\textsc{iv}\fi}
\newcommand{  \CIV      }{\ifmmode {\rm C}\,\textsc{iv}\,\lambda1549 \else C\,\textsc{iv}\,$\lambda1549$\fi}
\newcommand{  \NIIopt   }{\ifmmode \left[{\rm N}\,\textsc{ii}\right]\,\lambda6584 \else [N\,\textsc{ii}]\,$\lambda6584$\fi}
\newcommand{  \nii      }{\ifmmode \left[{\rm N}\,\textsc{ii}\right]  \else [N\,\textsc{ii}]\fi}
\newcommand{\oiii}{\ifmmode \left[{\rm O}\,\textsc{iii}\right] \else [O\,{\sc iii}]\fi}
\newcommand{\OIII}{\ifmmode \left[{\rm O}\,\textsc{iii}\right]\,\lambda5007 \else [O\,{\sc iii}]\,$\lambda5007$\fi}
\newcommand{  \mgii     }{\ifmmode {\rm Mg}\,\textsc{ii} \else Mg\,\textsc{ii}\fi}
\newcommand{  \MgII     }{\ifmmode {\rm Mg}\,\textsc{ii}\,\lambda2798 \else Mg\,\textsc{ii}\,$\lambda2798$\fi}
\newcommand{ \fwhm  }{\ifmmode {\rm FWHM} \else FWHM\fi} 
\newcommand{ \fwhb  }{\ifmmode {\rm FWHM}\left(\hb\right) \else FWHM(\hb)\fi}
\newcommand{ \ewhb  }{\ifmmode {\rm EW}\left(\hb\right) \else EW(\hb)\fi}
\newcommand{ \fwha  }{\ifmmode {\rm FWHM}\left(\ha\right) \else FWHM(\ha)\fi}
\newcommand{ \ewha  }{\ifmmode {\rm EW}\left(\ha\right) \else EW(\ha)\fi}
\newcommand{  \lamLlam  }{\ifmmode \lambda L_{\lambda} \else $\lambda L_{\lambda}$\fi}
\newcommand{  \nuLnu    }{\ifmmode \nu L_{\nu} \else $\nu L_{\nu}$\fi}
\newcommand{  \Lop      }{\ifmmode L_{5100} \else $L_{5100}$\fi}
\newcommand{  \Luv      }{\ifmmode L_{1450} \else $L_{1450}$\fi}
\newcommand{\fbol}{\ifmmode f_{\rm bol} \else $f_{\rm bol}$\fi}
\newcommand{\fbolopt}{\ifmmode f_{\rm bol}\left(5100{\rm \AA}\right) \else $f_{\rm bol}\left(5100{\rm \AA}\right)$\fi}
\newcommand{  \mbh      }{\ifmmode M_{\rm BH} \else $M_{\rm BH}$\fi}
\newcommand{  \lledd    }{\ifmmode L/L_{\rm Edd} \else $L/L_{\rm Edd}$\fi}
\newcommand{  \Lbol     }{\ifmmode L_{\rm bol} \else $L_{\rm bol}$\fi}
\newcommand  {\RBLR}        {\hbox{$ {R_{\rm BLR}} $}}
\newcommand{  \hst     }  {{\it HST}}
\newcommand{  \chandra }  {{\it Chandra}}
\newcommand{  \xmm     }  {{\it XMM-Newton}}
\newcommand{  \XMM     }  {{\it XMM-Newton}}

\newcommand{  \swift     }  {{\it Swift}}
\newcommand{  \nicer     }  {{\it NICER}}
\newcommand{  \NH       }{\ifmmode N_{\rm H} \else $N_{\rm H}$\fi}     

\newcommand{\mysobjat}{AT\,2018zf}
\newcommand{\mysobjas}{ASASSN-18el}
\newcommand{\mysobjagn}{1ES\,1927+654}
\newcommand{\mysobj}{\mysobjagn}
\newcommand{\zpaper}{$z=0.019422$}

\newcommand{\Nmonths}{11} 

\received{March 25, 2019}
\revised{June 25, 2019}
\accepted{July 13, 2019}
\submitjournal{ApJ}

\usepackage[normalem]{ulem}

\shorttitle{1ES 1927+654 as a changing look AGN}
\shortauthors{Trakhtenbrot et al.}

\begin{document}

\title{1ES 1927+654: An AGN Caught Changing Look on a Timescale of Months}

\correspondingauthor{Benny Trakhtenbrot}
\email{benny@astro.tau.ac.il}

\author[0000-0002-3683-7297]{Benny Trakhtenbrot}
\affil{School of Physics and Astronomy, Tel Aviv University, Tel Aviv 69978, Israel}

\author[0000-0001-7090-4898]{Iair Arcavi}
\affil{School of Physics and Astronomy, Tel Aviv University, Tel Aviv 69978, Israel}

\author[0000-0003-3422-2202]{Chelsea L. MacLeod}
\affil{Center for Astrophysics \textbar{} Harvard \& Smithsonian, 60 Garden Street, Cambridge, MA 02138-1516, USA}

\author[0000-0001-5231-2645]{Claudio Ricci}
\affil{N\'ucleo de Astronom\'ia de la Facultad de Ingenier\'ia, Universidad Diego Portales, Av. Ej\'ercito Libertador 441, Santiago, Chile}
\affil{Kavli Institute for Astronomy and Astrophysics, Peking University, Beijing 100871, China}

\author[0000-0003-0172-0854]{Erin Kara}
\affil{Joint Space-Science Institute, University of Maryland, College Park, MD 20742, USA}
\affil{Astrophysics Science Division, NASA Goddard Space Flight Center, 8800 Greenbelt Road, Greenbelt, MD 20771, USA}

\author[0000-0002-9154-3136]{Melissa L. Graham}
\affil{Department of Astronomy, University of Washington, Box 351580, U.W., Seattle, WA 98195-1580, USA}

\author[0000-0003-2686-9241]{Daniel Stern}
\affil{Jet Propulsion Laboratory, California Institute of Technology, 4800 Oak Grove Drive, MS 169-224, Pasadena, CA 91109, USA}

\author[0000-0003-2992-8024]{Fiona A. Harrison}
\affil{Cahill Center for Astronomy and Astrophysics, California Institute of Technology, 1200 E. California Blvd, Pasadena, CA 91125, USA}
%


\author{Jamison Burke}
\affil{Las Cumbres Observatory, 6740 Cortona Drive, Suite 102, Goleta, CA 93117-5575, USA}
\affil{Department of Physics, University of California, Santa Barbara, CA 93106-9530, USA}

\author[0000-0002-1125-9187]{Daichi Hiramatsu}
\affil{Las Cumbres Observatory, 6740 Cortona Drive, Suite 102, Goleta, CA 93117-5575, USA}
\affil{Department of Physics, University of California, Santa Barbara, CA 93106-9530, USA}

\author[0000-0002-0832-2974]{Griffin Hosseinzadeh}
\affil{Center for Astrophysics \textbar{} Harvard \& Smithsonian, 60 Garden Street, Cambridge, MA 02138-1516, USA}

\author[0000-0003-4253-656X]{D. Andrew Howell}
\affil{Las Cumbres Observatory, 6740 Cortona Drive, Suite 102, Goleta, CA 93117-5575, USA}
\affil{Department of Physics, University of California, Santa Barbara, CA 93106-9530, USA}


\author[0000-0002-8229-1731]{Stephen J. Smartt}
\affil{Astrophysics Research Centre, School of Mathematics and Physics, Queens University Belfast, Belfast BT7 1NN, UK}

\author[0000-0002-4410-5387]{Armin Rest}
\affil{Space Telescope Science Institute, 3700 San Martin Drive, Baltimore, MD 21218, USA}
\affil{Department of Physics \& Astronomy, Johns Hopkins University, 3400 North Charles Street, Baltimore, MD 21218, USA}
%


\author[0000-0003-0943-0026]{Jose L. Prieto}
\affil{N\'ucleo de Astronom\'ia de la Facultad de Ingenier\'ia, Universidad Diego Portales, Av. Ej\'ercito Libertador 441, Santiago, Chile}
\affil{Millennium Institute of Astrophysics, Santiago, Chile}

\author[0000-0003-4631-1149]{Benjamin J. Shappee}
\affil{Institute for Astronomy, University of Hawai’i, 2680 Woodlawn Drive, Honolulu, HI 96822, USA}

\author[0000-0001-9206-3460]{Thomas W.-S. Holoien}
\affil{The Observatories of the Carnegie Institution for Science, 813 Santa Barbara Street, Pasadena, CA 91101, USA}

\author[0000-0001-7485-3020]{David Bersier}
\affil{Astrophysics Research Institute, Liverpool John Moores University, 146 Brownlow Hill, Liverpool L3 5RF, UK}


\author[0000-0003-3460-0103]{Alexei V. Filippenko}
\affil{Department of Astronomy, University of California, Berkeley, CA 94720-3411, USA}
\affil{Miller Senior Fellow, Miller Institute for Basic Research in Science,
University of California, Berkeley, CA 94720, USA}

\author{Thomas G. Brink} 
\affil{Department of Astronomy, University of California, Berkeley, CA 94720-3411, USA}

\author[0000-0002-2636-6508]{WeiKang Zheng}
\affil{Department of Astronomy, University of California, Berkeley, CA 94720-3411, USA}
%


\author{Ruancun Li}
\affil{Kavli Institute for Astronomy and Astrophysics, Peking University, Beijing 100871, China}

\author[0000-0003-4815-0481]{Ronald A. Remillard}
\affil{MIT Kavli Institute for Astrophysics and Space Research, 70 Vassar Street, Cambridge, MA 02139, USA}

\author[0000-0002-1661-4029]{Michael Loewenstein}
\affil{Astrophysics Science Division, NASA Goddard Space Flight Center, 8800 Greenbelt Road, Greenbelt, MD 20771, USA}
\affil{Department of Astronomy, University of Maryland, College Park, MD 20742, USA}

\begin{abstract}
We study the sudden optical and ultraviolet (UV) brightening of 1ES\,1927+654, which until now was known as a narrow-line active galactic nucleus (AGN). 
1ES\,1927+654 was part of the small and peculiar class of ``true Type-2'' AGN, which lack broad emission lines and line-of-sight obscuration.
Our high-cadence spectroscopic monitoring captures the appearance of a blue, featureless continuum, followed several weeks later by the appearance of broad Balmer emission lines.
This timescale is generally consistent with the expected light travel time between the central engine and the broad-line emission region in (persistent) broad-line AGN.
{\it Hubble Space Telescope} spectroscopy reveals no evidence for broad UV emission lines (e.g., C\,\textsc{iv}\,$\lambda1549$, C\,\textsc{iii}]\,$\lambda1909$, Mg\,\textsc{ii}\,$\lambda2798$), probably owing to dust in the broad-line emission region. 
To the best of our knowledge, this is the first case where the lag between the change in continuum and in broad-line emission of a ``changing-look'' AGN has been temporally resolved.
The nature and timescales of the photometric and spectral evolution disfavor both a change in line-of-sight obscuration and a change of the overall rate of gas inflow as driving the drastic spectral transformations seen in this AGN. 
Although the peak luminosity and timescales are consistent with those of tidal disruption events seen in inactive galaxies, the spectral properties are not.
The X-ray emission displays a markedly different behavior, with frequent flares on timescales of hours to days, and will be presented in a companion publication.
\end{abstract}


\section{Introduction}
\label{sec:intro}

Large-scale time-domain surveys have allowed the recent identification of new types of extreme variability among both dormant and active supermassive black holes (SMBHs).
Among these, so-called ``changing look active galactic nuclei'' (CL-AGN, hereafter) are characterized by a switch between spectral states that are dominated by an AGN-like, power-law optical/ultraviolet (UV) continuum and/or strong broad emission lines ($>$1000\,\kms; i.e., Type-1 AGN), and those dominated by stellar continuum emission (from the host galaxy), and only narrow forbidden and permitted transitions from low- and high-ionization species (i.e., Type-2 AGN).
Many recent studies have identified a growing numbers of such cases \cite[e.g.,][and references therein]{Denney2014_Mrk590,Shappee2014_ASASSN_N2617,LaMassa2015_changing,MacLeod2016_CLAGN,McElroy2016_Mrk1018_CL,Runnoe2016_changing,Ross2018_CLAGN_model,Stern2018_CLAGN_WISE,Yang2018_CLAGN,Wang2018_CLAGN}, 
as well as cases of drastic changes to line-of-sight obscuration, identified in the X-rays \cite[e.g.,][]{Matt2003_CLAGN_Xrays,Piconcelli2007_CLAGN,Ricci2016_IC751_CLAGN}.

The drastic changes seen in CL-AGN challenge the simplest form of the AGN unification framework \cite[e.g.,][]{Antonucci1993,UrryPadovani1995_rev}, where the two main spectral types are explained by orientation, and where Type-2 AGN are the dust-obscured counterparts to broad-line, Type-1 AGN. 
A transient event in which a dusty cloud is going into (or out of) the line of sight is a rather unlikely explanation for CL-AGN, as the cloud would have to cover a significant fraction of the broad-line emission region (i.e., BLR), with sizes $ \gg 10^{16}\,{\rm cm}$, to account for the Type-2 appearance.
More likely explanations are that CL-AGN are driven by dramatic changes to the accretion rate onto the SMBH, or indeed the reformation or truncation of a radiatively-efficient accretion flow;
and/or 
by a sudden change to the amount of dense circumnuclear gas, that would give rise to broad emission lines (that is, BLR gas).
The timescales relevant for the CL-AGN reported up until now, of order several years, are far shorter than what is expected for global accretion rate changes in optically thick, geometrically thin accretion disks (see  discussion in, e.g.,  \citealt{Lawrence2018_viscosity_crisis,Stern2018_CLAGN_WISE}, as well as the historical commentary given by \citealt{Antonucci2018_viscosity_comment}).
However, the sparse spectroscopic sampling of these CL-AGN -- with ``before'' and ``after'' spectra usually taken years apart -- has so far prohibited a direct and clear-cut test of these scenarios.

\mysobjagn\ is a known redshift \zpaper\ AGN, based on its X-ray emission, identified with the {\it Einstein} and {\it ROSAT} satellites. 
The strong, narrow forbidden lines of high-ionization species (\OIII, \NIIopt), observed in optical spectra taken in 2001 June, further classify \mysobjagn\ as a Type-2 AGN \cite[e.g.,][]{Boller2003_1ES1927,Tran2011_trueSy2}.
Several studies of \mysobjagn\ have suggested that it challenges the AGN unification framework, by pointing out the very small amount of obscuring material along the line of sight 
(i.e., a hydrogen column density of $\log\left[\NH/\cmii\right] \lesssim 21.1$; see \citealt{Gallo2013_1ES1927}) and the lack of broad lines seen in polarized light \citep{Tran2011_trueSy2}.
Thus, \mysobj\ was proposed to be a prime example for the rare class of ``true'' or ``naked'' Type-2 AGN, intrinsically lacking the denser gas that gives rise to the BLR, and/or the photoionizing continuum radiation that drives the broad-line emission \cite[see the very recent study by][]{Bianchi2019_NGC3147_BLR}.

Here we report a dramatic optical and UV brightening of \mysobjagn, accompanied by the appearance of prominent broad emission lines with a time delay consistent with the expected size of a BLR.
%

\begin{figure*}
\centering
\includegraphics[angle=-00,width=0.9\textwidth]{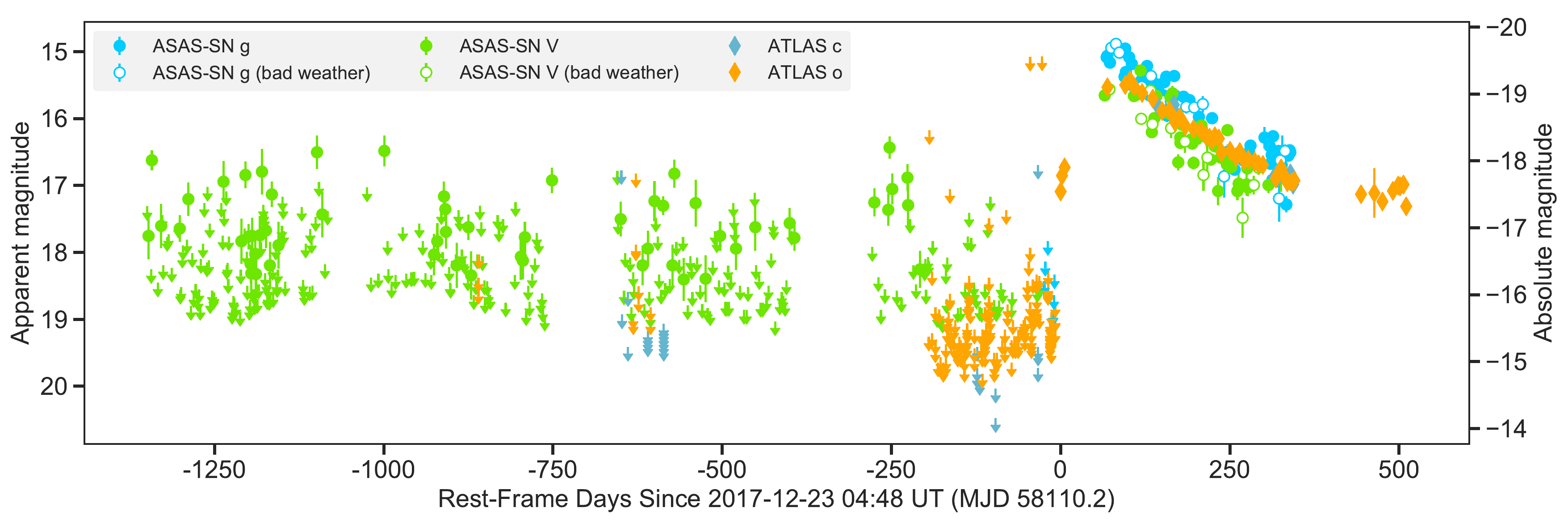} 
\caption{
Long-term optical light curve of \mysobjagn\ showing no activity at the level of the most recent flare (2017 December 23) in the last $\sim$4 yr. 
Arrows denote $3\sigma$ nondetection upper limits. 
All detections are binned nightly, 
and we separate ASAS-SN detections where clouds are seen in visual inspection of the images (marked by empty symbols).
Error bars denote $1\sigma$ uncertainties and are sometimes smaller than the symbols. 
All magnitudes are corrected for Milky Way extinction.
}
\label{fig:long_term_LC}
\end{figure*} 

\begin{figure*}
\centering
\includegraphics[angle=-00,width=0.8\textwidth]{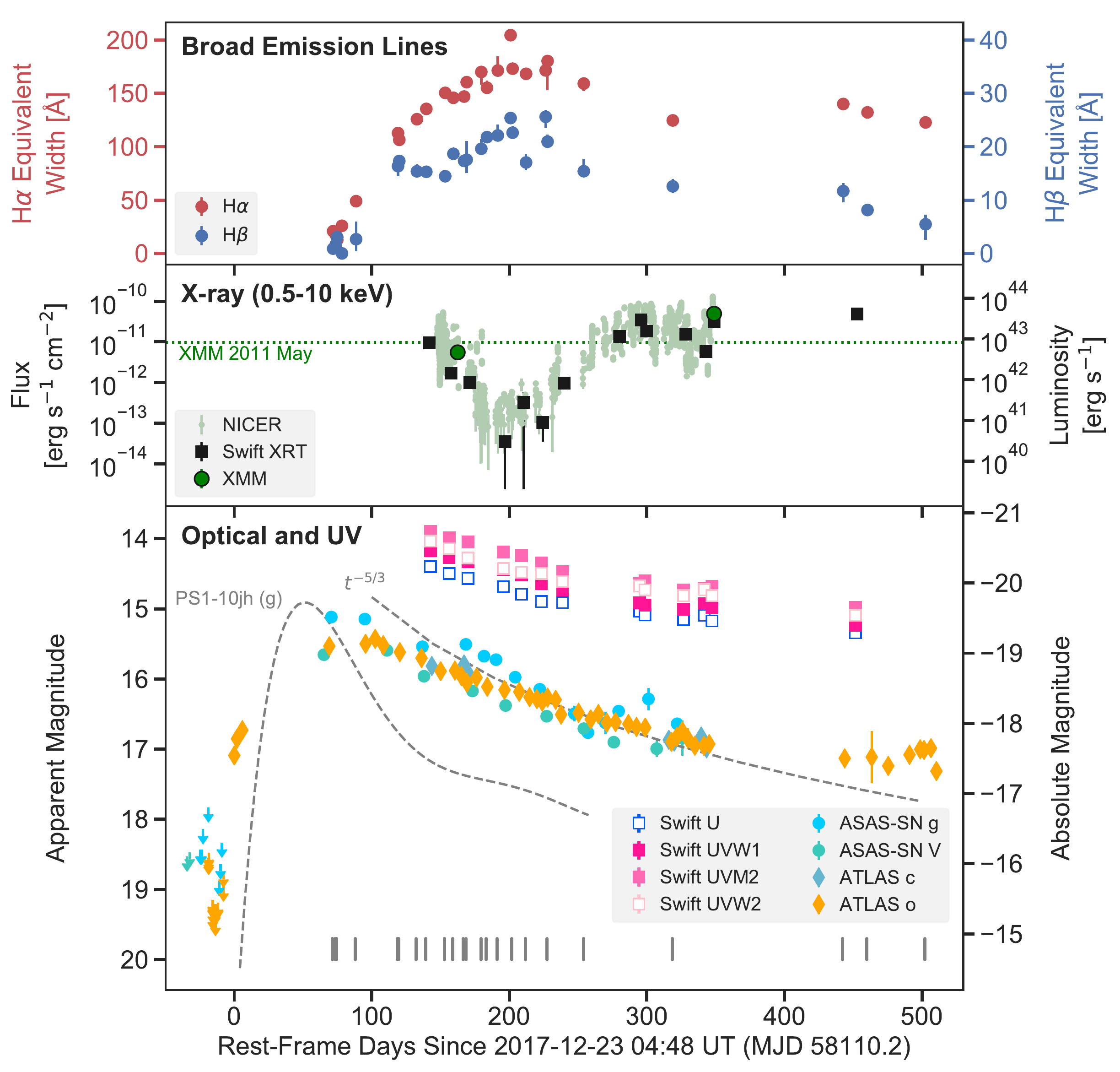} 
\caption{
Light curve and equivalent-width evolution of \mysobjagn\ during the flare.
{\it Bottom:} Optical and UV light curves, obtained with ATLAS, ASAS-SN, and \swift/UVOT.
Arrows denote $3\sigma$ nondetection upper limits. 
Empty symbols denote non-host-subtracted data.
The ATLAS data are binned nightly up to peak, and then binned every two nights. 
The ASAS-SN data are binned weekly. 
Error bars denote $1\sigma$ uncertainties.
All magnitudes are corrected for Milky Way extinction.
Vertical lines near the bottom of the panel denote epochs of optical spectra.
The dashed gray lines trace a simple $\propto t^{-5/3}$ power law and the $g$-band light curve of the TDE PS1-10jh \cite[matched in absolute magnitude;][]{Gezari2012_TDE}.
{\it Middle:} X-ray light curve obtained with \swift/XRT, \nicer, and \XMM. 
The spread in \nicer\ measurements is real, tracing dramatic and fast variability (order of magnitude changes seen day-to-day). 
The horizontal dotted line marks the 2011 May archival flux level in the \xmm\ 0.3--10 \kev\ band \citep{Gallo2013_1ES1927}.
{\it Top:} Equivalent-width evolution of the broad components of the \Hbeta\ and \Halpha\ emission lines.
}
\label{fig:Swift_LC}
\end{figure*} 

%
\section{Observations}
\label{sec:obs}

\subsection{Photometry}
\label{subsec:photo}

An increase in optical flux from \mysobjagn\ was discovered on 2018 March 3 (UT dates are used throughout; \citealt{Nicholls2018_ATel_1ES1927,Stanek2018_AS_report_18zf}) and announced by the All-Sky Automated Survey for Supernovae \cite[ASAS-SN;][]{Shappee2014_ASASSN_N2617}, with a reported $V$-band host-subtracted magnitude of 15.3.
The transient event was reported as \mysobjas\ and given the name \mysobjat\ by the transient name server\footnote{\url{http://wis-tns.weizmann.ac.il/object/2018zf}}.
Pre-discovery detections from 2017 December 23 were recovered by the Asteroid Terrestrial-impact Last Alert System \citep[ATLAS;][]{Tonry2018_ATLAS} and designated as ATLAS18mgv.
In what follows, we adopt this earlier ATLAS pre-discovery measurement as the detection date.

ASAS-SN images were processed by the fully automatic ASAS-SN pipeline \citep{Shappee2014_ASASSN_N2617} using the ISIS image subtraction package \citep{AlardLupton1998,Alard2000_img_sub}. 
We performed aperture photometry on the subtracted, stacked images in order to increase the signal-to-noise ratio ($S/N$) of the detections, 
and calibrated the results using the AAVSO Photometric All-sky Survey \cite[APASS;][]{Henden2016_APASS}.

Forced photometry was run on all reference-subtracted ATLAS images to produce magnitudes and $3\sigma$ upper limits, using automated point-spread-function fitting, as documented by Tonry et al. (\citeyear{Tonry2018_ATLAS}; all ATLAS magnitudes are in the AB system, \citealt{OkeGunn1983_AB_mags}). 
In a large number of images, there were flux artifacts at the transient position, due to the bright host and imperfect subtraction.
These clearly spurious measurements were excised from the ATLAS light curve.

All optical magnitudes, as well as UV magnitudes and optical spectra (see below), were corrected for Milky Way extinction using $E(B-V)=0.077$ mag \citep{SchlaflyFinkbeiner2011_reddening}\footnote{Retrieved via the NASA/IPAC Extragalactic Database (NED): \url{http://ned.ipac.caltech.edu/}}, the \cite{Cardelli1989} extinction law, and $R_V=3.1$.

Following the sudden optical flux increase, we initiated a near-UV (NUV) and X-ray monitoring campaign using the Neil Gehrels \swift\ Observatory (PI: I.\ Arcavi). 
UVOT photometry was extracted with standard HEASARC functions using a 5\arcsec-radius aperture for the object and for the sky region. 
The host flux in the NUV $UVW1$ and $UVM2$ bands was determined from \XMM/OM data obtained in 2011 May 20 \cite[][AB magnitudes of 18.14 and 18.65, respectively, using the same aperture]{Gallo2013_1ES1927}, and subtracted from our new UVOT measurements.
The X-ray analysis is discussed in detail in a companion paper (Ricci et al., in prep.).
The final light curves are presented in Figures \ref{fig:long_term_LC} and \ref{fig:Swift_LC}.

\subsection{Spectroscopy}
\label{subsec:spec}

\begin{figure*}
\centering
\includegraphics[angle=-00,width=0.8\textwidth]{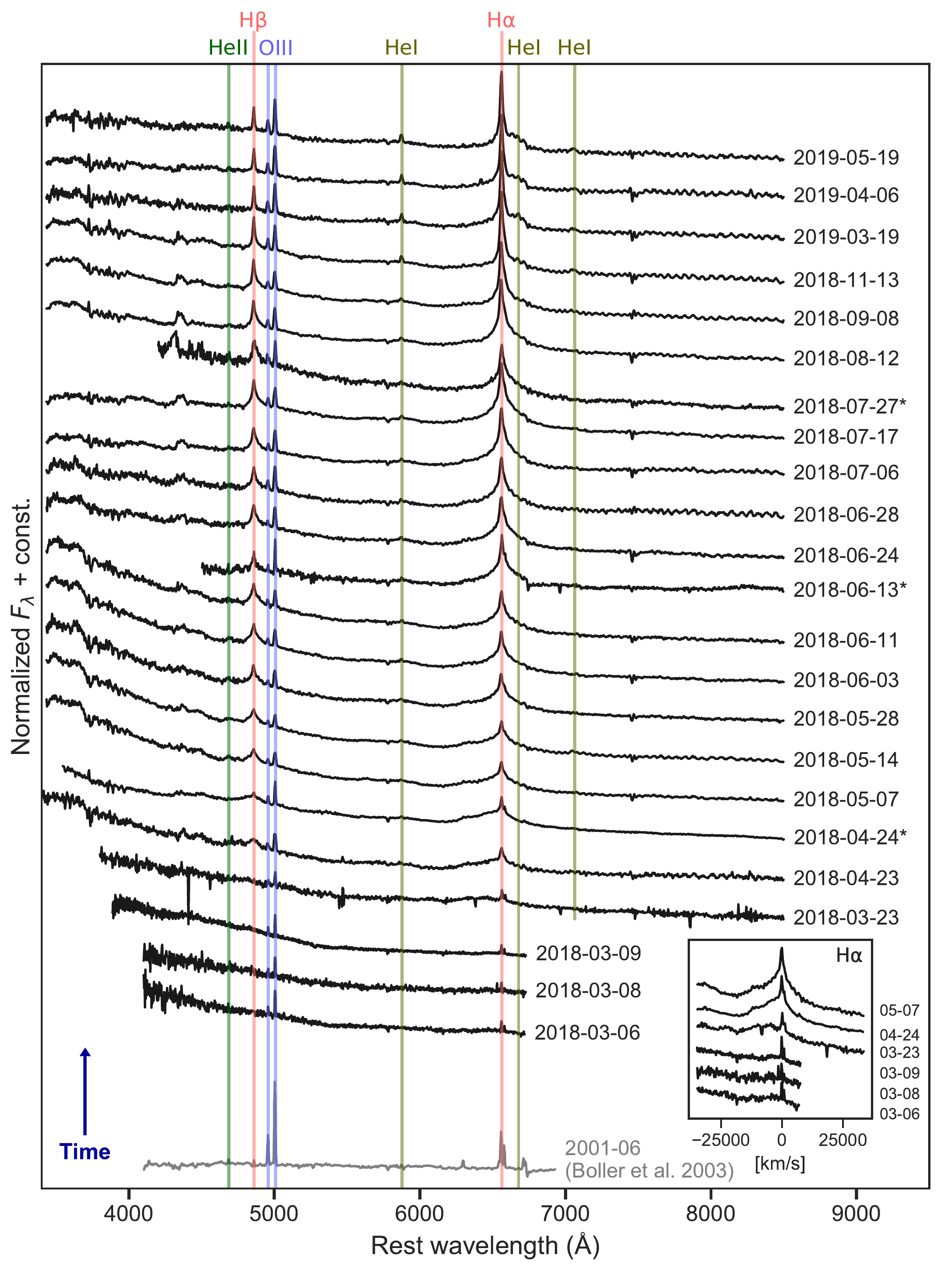} 
\caption{
Optical spectra of \mysobj\ showing the transition from a narrow-line (Type-2) AGN (bottom; from \citealt{Boller2003_1ES1927}) to a broad-line (Type-1) AGN with an intermediate stage of a blue-continuum-dominated emission. 
The appearance of the broad lines is constrained to a timescale of several weeks (between the 2018 March 6 and April 23 spectra), although there is some evidence for a weak, broad, and blueshifted \Halpha\ line in our first spectra, as shown in the inset.
All spectra are plotted without Milky Way extinction corrections.
Dates marked with $*$ indicate spectra taken under different weather and/or instrument conditions compared to a standard star, thus affecting the validity of their  continuum shape.
}
\label{fig:spec_monitoring}
\end{figure*}

We obtained optical spectra of \mysobjagn\ with
the OSMOS spectrograph mounted on the Hiltner 2.4~m Telescope at the MDM Observatory \citep{Martini2011_OSMOS}; 
the SPRAT spectrograph mounted on the 2~m Liverpool Telescope at 
the	Roque de los Muchachos Observatory \citep{Piascik2014_SPRAT};
the Kast spectrograph mounted on the 3~m Shane Telescope at 
the	Lick Observatory \citep{MillerStone1994_Lick_Kast};
the FLOYDS spectrograph mounted on the 2~m Faulkes Telescope North at Haleakala, Hawaii \cite[part of the Las Cumbres Observatory network;][]{Brown2013_LCOGT};
the Dual Imaging Spectrograph mounted on the Apache Point Observatory 3.5~m telescope; 
the Low Resolution Imaging Spectrograph (LRIS) mounted on the Keck-I telescope at Maunakea  \citep{Oke1995_LRIS};
and the Double Beam Spectrograph (DBSP) mounted on the Palomar Hale 5~m telescope \citep{OkeGunn1982_DBSP}.
All optical spectra were reduced following standard procedures, and scaled to match the ATLAS {\it orange}-band measurements, linearly interpolating between the nearest relevant ATLAS visits.
Whenever possible, the spectra were taken with slits rotated to the parallactic angle, to minimize chromatic losses and spectral distortions \citep{Filippenko1982_refraction}.
The sequence of optical spectra is shown in Figure \autoref{fig:spec_monitoring}.

\begin{figure*}
\centering
\includegraphics[trim={6.5cm 0 1cm 0},clip,width=0.975\textwidth]
{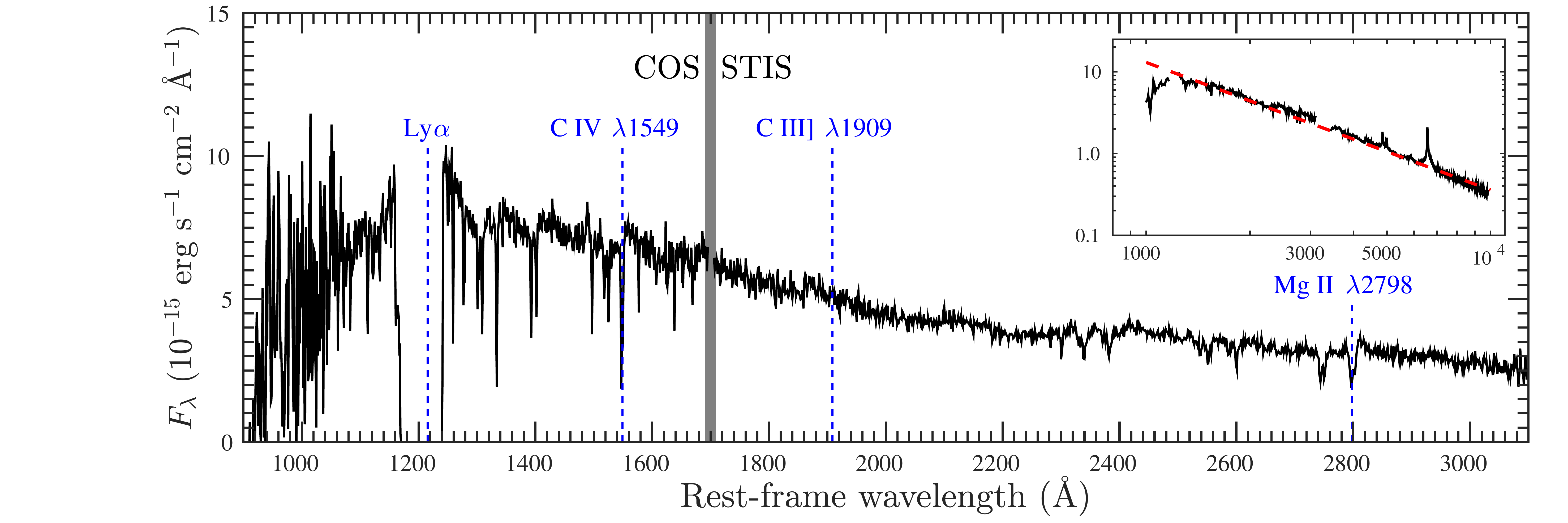}
\caption{
A combined \hst\ UV spectrum of \mysobjagn, obtained on 2018 August 28 with the COS (shortward of rest-frame wavelength of $\lambda_{\rm rest}=1700$ \AA) and STIS (longward of 1700 \AA) instruments.
The vertical lines mark the expected locations of the \CIV, \CIII, and \MgII\ transitions, commonly seen as prominent broad ($>1000\,\kms$) emission lines in broad-line AGN. 
The absence of such broad lines is consistent with the presence of dust in the BLR.
The inset compares the UV-optical spectral energyy distribution (SED), which includes the \hst\ spectra and the 2018 Sep. 8 Las Cumbres spectrum (all median-binned to 7 \AA), with a typical quasar power-law SED of the form $f_{\nu} \propto \nu^{-0.44}$ \citep{VandenBerk2001}.
This agreement indicates a mostly dust-free line-of-sight toward the continuum source.
} 
\label{fig:spec_uv_hst}
\end{figure*}

We obtained far-UV (FUV) and NUV  spectra of \mysobjagn\ using the COS and STIS instruments, respectively, on board the {\it Hubble Space Telescope} (\hst; Program ID 15604, PI: C.L.\ MacLeod).
Each spectrum was obtained through single-orbit exposures of 2866 and 2740 s, respectively, on August 28, and reduced following standard \hst\ procedures.
The two flux-calibrated \hst\ spectra are shown in \autoref{fig:spec_uv_hst}.

A log of all our spectral observations of \mysobj\ is detailed in \autoref{tab:obs_log_spec} and all our spectra are available for download from the Weizmann Interactive Supernova Data Repository (WISeREP)\footnote{\url{http://wiserep.weizmann.ac.il/}}.

We also obtained optical spectra of the four nearest (projected)  neighboring sources of \mysobjagn\ (sources within ${\sim}3{-}12\arcsec$, numbered 2, 3, 5, and 6 in Fig.~9 of \citealt{Boller2003_1ES1927}), using the FLOYDS spectrograph on the Las Cumbres Observatory 2~m Faulkes Telescope North. 
We find that all sources are Milky Way stars, which are not expected to contribute to the significant activity detected toward \mysobjagn.\footnote{None of these stars is detected in the \chandra\ data of \cite{Boller2003_1ES1927}.}

\begin{deluxetable*}{lccllc}
\tablecaption{Log of spectroscopic observations \label{tab:obs_log_spec}}
\tablecolumns{6}
\tablewidth{0pt}
\tablehead{
\colhead{Date\tablenotemark{a}} &
\colhead{MJD start} &
\colhead{$\Delta t_{\rm rf}$ \tablenotemark{b} } & \colhead{Telescope} & \colhead{Inst.} & \colhead{Exp.} \\
\colhead{} & \colhead{time\tablenotemark{a} (d)} & \colhead{(d)} & \colhead{} & \colhead{} 
& \colhead{time (s)}}
\startdata
2018-03-06     & 58183.5 &  72 & MDM Hiltner 2.4 m & OSMOS   & 600\\
2018-03-08     & 58185.5 &  74 & MDM Hiltner 2.4 m & OSMOS   & 1800\\
2018-03-09     & 58186.5 &  75 & MDM Hiltner 2.4 m & OSMOS   & 2700 \\
2018-03-13\tablenotemark{c} & 58190.0 &  79 & Liverpool Telescope 2 m & SPRAT   & 600\\
2018-03-23$^*$ & 58200.5 &  89 & APO 3.5 m         & DIS     & 300 \\
2018-04-23     & 58231.6 & 119 & Las Cumbres 2 m   & FLOYDS  & 1800\\
2018-04-24     & 58232.5 & 120 & Lick Shane 3 m         & Kast    & ${\sim}1500^\dagger$ \\
2018-05-07     & 58245.6 & 133 & Las Cumbres 2 m   & FLOYDS  & 1800\\
2018-05-14     & 58252.5 & 140 & Las Cumbres 2 m   & FLOYDS  & 1800\\
2018-05-28     & 58266.5 & 153 & Las Cumbres 2 m   & FLOYDS  & 1800\\
2018-06-03     & 58272.6 & 159 & Las Cumbres 2 m   & FLOYDS  & 1800\\
2018-06-11     & 58280.6 & 167 & Las Cumbres 2 m   & FLOYDS  & 1800\\
2018-06-13$^*$ & 58282.5 & 169 & APO 3.5 m         & DIS     &  600\\
2018-06-24     & 58293.3 & 180 & Las Cumbres 2 m   & FLOYDS  & 1800\\
2018-06-28     & 58297.5 & 184 & Las Cumbres 2 m   & FLOYDS  & 3600\\
2018-07-06     & 58305.5 & 192 & Las Cumbres 2 m   & FLOYDS  & 3600\\
2018-07-16     & 58315.5 & 201 & Keck 10 m         &  LRIS   &  900\\
2018-07-17     & 58316.5 & 202 & Las Cumbres 2 m   & FLOYDS  & 3600\\
2018-07-27     & 58326.5 & 212 & Las Cumbres 2 m   & FLOYDS  & 3600\\
2018-08-11     & 58341.0 & 227 & Palomar Hale 5 m  & DBSP    & 1200\\
2018-08-12     & 58342.4 & 228 & Las Cumbres 2 m   & FLOYDS  & 3600\\
2018-09-08     & 58369.3 & 254 & Las Cumbres 2 m   & FLOYDS  & 3600\\
2018-11-13     & 58435.2 & 319 & Las Cumbres 2 m   & FLOYDS  & 3600\\
2019-03-19     & 58561.5 & 443 & Las Cumbres 2 m   & FLOYDS  & 2700 \\
2019-04-06     & 58579.5 & 461 & Las Cumbres 2 m   & FLOYDS  & 2700 \\
2019-05-19     & 58622.6 & 503 & Las Cumbres 2 m   & FLOYDS  & 2700 \\
\hline													
2018-08-28     & 58358.4 & 244 & {\it HST} 2.4 m   & COS     & 2866\\ 
2018-08-28     & 58358.5 & 244 & {\it HST} 2.4 m   & STIS    & 2740   
\enddata
\tablenotetext{a}{At exposure start, UT. Dates marked with $*$ indicate spectra taken under different weather and/or instrument conditions compared to a standard star, thus affecting the validity of their  continuum shape.}
\tablenotetext{b}{Time since ATLAS first detection of the optical flare, in rounded rest-frame days.}
\tablenotetext{c}{A low-resolution spectrum, not shown in \autoref{fig:spec_monitoring}.}
\tablenotetext{\dagger}{~The Shane/Kast exposure times were $3 \times 500$~s for the red arm, and $1 \times 1560$~s for the blue.}
\end{deluxetable*}

\bigskip
\section{Analysis}
\label{sec:analysis}

The NUV emission from \mysobjagn, as measured from our first \swift\ observations, has increased by a factor of about 40 compared with the 2011 May \xmm\ measurement \citep{Gallo2013_1ES1927}. 
The XRT data (\autoref{fig:Swift_LC}), together with new data we obtained with \xmm\ and the {\it Neutron star Interior Composition ExploreR} \cite[{\it NICER};][]{Gendreau2012_NICER}, show more complex behavior compared to the optical and UV light curve, including significant variability on timescales of hours seen in the \nicer\ rapid monitoring. 
These X-ray data are analyzed in detail in a companion paper (Ricci et al., in prep.).

The first post-flare optical spectra are dominated by blue continuum emission, as well as several narrow emission lines, including \Hbeta, \OIII, \Halpha, and \NIIopt. 
There is evidence for a weak, broad, and blueshifted \Halpha\ emission feature in these first post-flare spectra (see inset of \autoref{fig:spec_monitoring}), however the limited spectral coverage does not allow for a robust measurement.
The narrow-line emission is consistent with gas photoionized by an AGN-like continuum, according to emission-line ratio diagnostics \cite[i.e., ``BPT diagnostics''; see, e.g.,][and references therein]{Kewley2006}, and is therefore consistent with previous studies of \mysobjagn. 
However, the blue continuum was not present in archival data \cite[][]{Boller2003_1ES1927,Tran2011_trueSy2}.

Strong, broad \halpha\ and \hbeta\ lines appeared on top of the blue continuum between 2018 March 6 and April 23 (\autoref{fig:spec_monitoring}). %
These lines remain strong during our entire spectroscopic monitoring campaign, lasting at least \Nmonths\ months after the optical flare detection (see top panel of \autoref{fig:Swift_LC}). 
To our knowledge this is the first time an AGN has been caught in the act of changing its type, with a blue continuum seen to appear before the broad lines, and the timescale for the broad-line appearance constrained to within about a month (i.e., between the 2018 March 23 and April 23 spectra).
The strength of the broad emission lines seems to evolve in a way that is generally similar to the UV/optical continuum emission (bottom panel of \autoref{fig:Swift_LC}), peaking with a delay of roughly 150 (rest-frame) days after the continuum peak, and then (slowly) declining.
The rise of the broad lines is, however, much slower than that of the continuum.

The FUV (\hst/COS) spectrum shows evidence for broad \Lya\ line emission, although the core of the line is missing owing to the physical gap in the COS detectors. 
Most notably, the \hst\ spectra do not show the prominent broad \CIV, \CIII, and \MgII\ emission lines, commonly seen in the spectra of ``normal,'' persistent broad-line AGN (quasars). 
The combined UV (\hst, 2018 Aug.~28) and optical (Las Cumbres, 2018 Sep.~8) post-flare spectrum of \mysobjagn\ is in excellent agreement with the typical spectral shape of such  broad-line AGN (i.e., $f_\nu \sim \nu^{-0.5}$, \citealt{VandenBerk2001}; see inset of \autoref{fig:spec_uv_hst}).
Moreover, the UV-optical SED -- as probed through the \swift/UVOT $UVM2-V$ color measurements -- shows very little evolution during our monitoring.
All this justifies some of the spectral analysis steps detailed below.

To measure the spectral features observed in our optical data, we decomposed the spectral regions surrounding the \hbeta, \mgii, and \civ\ lines, following the procedures presented by \cite{TrakhtNetzer2012_Mg2}. 
The region surrounding the \Halpha\ line was decomposed using a procedure that follows the one presented by \cite{Mejia2016_XS_MBH}.
The spectral models include a local (linear) continuum, a broadened iron emission template, single Gaussians for the narrow features of the \hbeta, \oiii, \halpha, and \nii\ lines, and two broad Gaussians for each of the broad emission lines.
Uncertainties on key  quantities were derived through a resampling procedure, fitting 100 realizations of each of the observed spectra assuming the observed noise, and eventually adopting the 16th and 84th quantiles as 1{$\sigma$} equivalent errors on the quantities in question.

The top panel of \autoref{fig:Swift_LC} shows the time evolution of the rest-frame equivalent widths (EW) of the broad \hbeta\ and \halpha\ lines.
There is a clear rise in the broad \ewha\ between 2018 March 9 and 23 (i.e., about 70--90 days after the optical transient detection), completing an order-of-magnitude increase to $\ewha\approx110$ \AA\ by April 23 (120 days after detection, in the rest frame). 
By this time the broad \hbeta\ line can be robustly identified, with $\ewhb=16\pm2$ \AA.
Upon first robust appearance of a broad \halpha\ line (in the March 23 spectrum), it is highly blueshifted, with $\Delta v \approx -5000\,\kms$ compared to the systemic redshift (and to the narrow \halpha\ component), and broad, with a full width at half-maximum intensity (FWHM) of $\fwha\approx18,000\,\kms$.
At later times the line emission remains roughly constant, with EWs in the range $\ewhb{\approx}15-27$ \AA\ and $\ewha{\approx}125-180$ \AA. 
The \halpha\ strength in \mysobj\ is rather typical of low-redshift broad-line AGN, with roughly half of such systems having $\ewha \lesssim 200$ \AA\ (see, e.g., the luminous AGN samples studied by \citealt{Shen_dr7_cat_2011} and \citealt{Koss2017_BASS_DR1}, and also \citealt{Stern2012_T1_lowz} and \citealt{Oh2015_hidden_BLAGN} for less-luminous systems).
The \Hbeta\ emission in \mysobj, on the other hand, is rather weak, compared with the $\sim$95\% of vigorously accreting broad-line, low-redshift AGN that have $\ewhb > 25$ \AA\ \cite[see also][]{TrakhtNetzer2012_Mg2}.

Throughout our spectroscopic monitoring, the broad \halpha\ to \hbeta\ flux ratio, which probes the Balmer decrement, was high: it reached $F(b {\rm H}\alpha) / F(b {\rm H}\beta) \gtrsim 8 $ about 150 days after the transient detection (in the rest frame), and further increased up to $\gtrsim15$, as the broad \hbeta\ line flux was decreasing. 
These line ratios are much higher than what is typically seen in broad-line AGN \cite[e.g.,][]{Dong2008_Balmer_decrement}, and are indicative of a highly dusty BLR. 
This can also account for the nondetection of UV line emission: our \hst\ spectra resulted in upper limits of ${\rm EW}(\civ)<1$ \AA\ and ${\rm EW}(\mgii)<5$ \AA, which are indeed much lower that what is usually found for broad-line AGN \cite[e.g.,][]{Baskin2005_CIV,Sulentic2007,Richards2011_CIV,Shen_dr7_cat_2011,Tang2012}.
The steep Balmer decrement suggests extinction levels of at least $A_{\rm line} \approx 6.5$ and 4.8 mag at the central wavelengths of the \civ\ and \mgii\ lines, respectively.\footnote{Assuming again the \citet{Cardelli1989} extinction law and $R_V=3.1$.}
If one would scale up our UV line EW upper limits based on these significant levels of extinction, then our data could be consistent with the distribution of broad UV emission line strength, as seen in normal broad-line AGN.
Recalling that the {\it continuum} emission does not show any evidence for significant dust extinction (inset of \autoref{fig:spec_uv_hst}; but see also \citealt{Baron2016_dust}), we conclude that a high dust content {\it within} the broad-line-emission region is consistent with all the UV-optical data in hand. 
This combination is quite rare in normal, persistent broad-line AGN \cite[e.g.,][]{Baron2016_dust}. 
However, it could perhaps be expected if the broad emission lines originate from gas in which the dust in the parts closer to the central engine were exposed to a recent ``flash'' of sublimating (UV) continuum radiation. 
We discuss this point further in \autoref{sec:discussion}.

To derive rough estimates of the size of the newly detected broad-emission-line region, \RBLR, we rely on $\RBLR-L$ relations derived from reverberation mapping campaigns.
As the broad-line emission is essentially driven by the ionizing (UV) radiation from the central engine, we can use the highest \swift-measured NUV monochromatic luminosity\footnote{We assume a cosmological model with $\Omega_{\Lambda}=0.7$, $\Omega_{\rm M}=0.3$, and $H_{0}=70\,\kms\,{\rm Mpc}^{-1}$.} of $\lamLlam({\rm NUV})\approx1.2\times10^{44}\,\ergs$ (in the $UVM2$ band, $\lambda_{\rm eff}=2262$ \AA), and the $\RBLR-\Luv$ relation of \cite{Kaspi2005}, to derive a BLR size of $\RBLR({\rm UV})\approx30$ light-days. %
This estimate is likely a lower limit, as our earliest \swift\ observation took place after the  optical peak (see \autoref{fig:Swift_LC}), and since the SED of the source is likely to further rise from the observed NUV band toward, and indeed beyond, the Lyman limit \cite[see][and references therein]{Shull2012}. 

Using instead the optical monochromatic luminosity at rest-frame 5100 \AA, \lamLlam(5100 \AA) (hereafter \Lop), measured from the first spectrum that robustly shows both broad Balmer emission lines (the 2018 April 23 spectrum), which is $\Lop=9.6\times10^{42}\,\ergs$, and the $\RBLR-\Lop$ prescription of \cite{Bentz2013_lowL_RL}, we derive $\RBLR \approx 10$ light-days.

The BLR sizes we obtain through $\RBLR-L$ relations, on the order of tens of light days, are consistent with the delay of 1--3 months between the UV/optical flux increase and the appearance of the broad lines, to within the uncertainties related with the $\RBLR-L$ relations, which are mostly due to systematics. 
Indeed, the source-to-source scatter around the $\RBLR-L$ relations is of order 0.2 dex  \cite[e.g.,][]{Bentz2013_lowL_RL}.

We further use the Keck/LRIS spectrum taken on 2018 July 16 to estimate key properties of the accreting SMBH powering \mysobjagn.
This higher resolution spectrum probes the broad Balmer lines about 12 weeks after their appearance.
The best-fit spectral model results in
a broad \hbeta\ line width of $\fwhb=3100^{+70}_{-80}\,\kms$. 
The (narrow) \OIII\ line peak indicates a redshift of 0.01942, which we adopt throughout.
Combining the aforementioned \Lop-based BLR size estimate of 10 light days with the broad \hbeta\ width, and a virial factor $f{=}1$, we obtain a virial (``single-epoch'') BH mass estimate of $\mbh \approx 1.9{\times}10^{7}\,\Msol$.
The uncertainties on such mass estimates are of order $0.3{-}0.5$ dex -- dominated by systematic uncertainties on the $\RBLR-\Lop$ relation and on the virial factor $f$ (see \citealt{Shen2013_rev} for a detailed discussion). 
Our \mbh\ estimate is in excellent agreement with the one reported by Tran \et (\citeyear{Tran2011_trueSy2};  $2.2\times10^{7}\,\Msol$), which was based on far less robust methods (i.e., the narrow \OIII\ line).

Using the aforementioned 2018 April 23 measurement of \Lop\ and a bolometric correction of $\fbolopt=9$ \cite[e.g.,][]{Kaspi2000,Runnoe2012}, we derive a bolometric luminosity of $\Lbol{=}8.6{\times}10^{43}\,\ergs$, close to the highest optical flux levels covered by our observations (about 120 rest-frame days after the transient discovery). 
Combining this with our \mbh\ estimate, we derive an Eddington ratio of $\lledd \equiv \fbolopt \times \Lop / (1.5\times10^{38}\times [\mbh/\Msol])  \approx 0.03$.
Adopting the same \mbh\ estimate and bolometric correction to all our spectroscopic data (i.e., \Lop\ measurements), we obtain Eddington ratios in the range $\lledd\approx0.008-0.03$.
If we instead use the highest measured NUV luminosity and a conservative NUV bolometric correction of $f_{\rm bol}({\rm NUV})=2$ \cite[or 4; see, e.g.,][]{Runnoe2012,TrakhtNetzer2012_Mg2,Netzer2016_herschel_hiz}, we obtain $\Lbol{\approx}2.3{\times}10^{44}\,\ergs$ (or $4.6{\times}10^{44}\,\ergs$), and Eddington ratios of order $\lledd \approx 0.1$ (or 0.2, respectively). 
All these estimates of \Lbol\ and \lledd\ are highly uncertain, as the (optical to X-ray) SED shape of \mysobjagn\ during this intensified UV/optical emission episode, and thus the bolometric luminosity and corrections, are likely very different from what is normally seen and/or assumed for broad-line AGN.
A detailed analysis of the full SED of \mysobjagn\ during the transient event presented here, as well as the implied \Lbol\ and \lledd, and a comparison to previous studies, will be presented in the companion paper (Li et al., in prep.).

\section{Discussion and Conclusion}
\label{sec:discussion}

Our data show that \mysobjagn\ experienced a dramatic increase in continuum UV/optical emission, forming a blue, AGN-like continuum, which was then followed by the appearance of prominent broad Balmer emission lines. 
While both the (optical) continuum and line emission rose within about a month, the appearance of the broad Balmer lines lagged the continuum rise by 1--3 months. 
The uncertainty in the time-lag determination is due to the uncertainty in the exact time the UV flux peaked, and the exact time the broad lines emerged. 
Still, to our knowledge, this is the first observation of a changing-look AGN (CL-AGN) where the lag between the change in continuum and in broad-line emission has been temporally resolved. 
Broad UV lines, some of which probe higher levels of ionization (e.g., \CIV), remain undetected.
The delay between the continuum and broad-line emission, if taken as the light travel time to the BLR, allows us to estimate a BLR radius (\RBLR) of roughly 1--3 light-months.
Moreover, our spectroscopic time series (\autoref{fig:spec_monitoring}) shows that the highest-velocity BLR gas was observed to respond first, followed later by a steady increase in the lower-velocity, core line emission.

These dramatic changes seen in \mysobj\ are unlikely to be driven by a change in the level of line-of-sight obscuration (i.e., a dusty cloud moving out of the line-of-sight), as in such a case one would expect no delay between the appearance of the two emission components.
Indeed, our rich collection of X-ray data (Ricci et al., in prep.) offers no evidence for a coherent and persistent X-ray spectral change that could straightforwardly be linked to a change in line-of-sight obscuration. 
We also recall that \mysobjagn\ was identified as one of the few ``true type-2'' AGN\footnote{It most obviously no longer fits into this class.}
 -- systems which show neither broad-line emission nor significant line-of-sight obscuration (which might have accounted for the lack of broad lines). 
Such systems are thought to have either a low content of broad-line emitting circumnuclear gas (i.e., no BLR gas) and/or insufficient levels of ionizing continuum emission.

All this suggests that the changes seen in \mysobj\ are most likely driven by a sudden change in the accretion flow onto the SMBH, resulting in increased UV/optical continuum emission which traveled to the BLR gas, where it was reprocessed and gave rise to broad-line emission. 

Here we briefly consider three possibilities for the nearly-cotemporaneous appearance of the blue continuum and broad-line emission. 
We then turn to briefly discuss some of the mechanisms that could be driving the entire (enhanced emission) event, including the tidal disruption of a star.

%
%
First, a fresh supply of (cold) gas may have reached the close vicinity of the SMBH. 
Some of the gas may have ended up being accreted onto the SMBH, thus illuminating circumnuclear gas on larger scales, some of which has properties consistent with that of the BLR seen in ``normal'' broad-line AGN. 
Given the \Lbol\ estimates available for \mysobjagn, the amount of gas that went through the AGN-like accretion flow during our monitoring is of order ${\lesssim}0.01\,\Msol$ (assuming a standard radiative efficiency of $\eta=0.1$), and the total mass in the broad emission line may be of order ${\sim}0.1\,\Msol$ \cite[e.g.,][]{Netzer2013_book}.
However, the X-ray and {\it narrow} line emission previously seen in \mysobjagn\ indicate that the SMBH has been vigorously accreting for over a decade (and perhaps through $\sim10^{3-4}$ yr).
Moreover, 
the month-long rise of the continuum and line emission is far faster than what is expected from a global rise in the accretion rate through an AGN-like disk, and/or from  heating/cooling fronts traveling within such a disk \mysobj\ \cite[see discussion in, e.g.,][]{LaMassa2015_changing,Lawrence2018_viscosity_crisis,Ross2018_CLAGN_model,Stern2018_CLAGN_WISE}.

%
%
Second, a drastic change in the accretion flow and in the related UV/optical continuum emission may have initiated broad-line emission in a pre-existing reservoir of BLR-like gas. 
This can be the result of enhanced accretion through a pre-existing accretion disk, although the observed timescales suggest that other, localized disturbances are much more likely (see the discussion in, e.g., \citealt{Stern2018_CLAGN_WISE,Ross2018_CLAGN_model}
and references therein).
One intriguing possibility is that a UV ``flash'' sublimated the dust in the inner regions of a dusty obscuring structure (i.e., a dusty ``torus''), making this inner-torus region appear as a de-facto newly-formed BLR \cite[see the review by][and references therein]{Netzer2015_torus_rev}.
Given the previous ``true Type-2'' classification of \mysobjagn, it is possible that the system is observed at a relatively small inclination angle (i.e., relatively ``face on''), and that the historically low UV emission allowed the torus to extend inwards. 
Indeed, the BLR gas may still have a significant dust content, which would account for the steep Balmer decrement and the nondetected broad UV emission lines. 
We recall again that the physical conditions and radiation transfer within the BLR gas in \mysobjagn\ may not be necessarily similar to those in normal broad-line AGN.

%
%
Finally, several models suggest that increased SMBH accretion may trigger disk  instabilities \cite[][]{Nicastro2003_true_type2} or indeed launch a disk-wind \cite[e.g.,][]{ElitzurHo2009_BLR,ElitzurNetzer2016_true_Sy2}, either of which would then be exposed to the incident ionizing radiation (from the inner disk) and thus be seen in broad-line emission. 
Some recent studies suggest observational evidence for a link between CL-AGN and such disk-winds \cite[e.g.,][]{Giustini2017_N2617,MacLeod2019_CLAGN}.
However, the relevance of this disk-wind scenario to the drastic changes seen in \mysobjagn\ is unclear, given 
(1) the absence of high-ionization, UV broad emission lines (\autoref{fig:spec_uv_hst}), which are expected to form in the inner region of a disk-wind \cite[e.g.,][and references therein]{Richards2011_CIV}; 
(2) the delay between the continuum and broad-line emission appearance is much longer than what is expected for a disk-wind;
and
(3) the lack of clear evidence for outflows in our UV and optical spectroscopy.
Moreover, the \cite{Nicastro2003_true_type2} scenario is rather unlikely, as it is expected to occur around a threshold Eddington rate of $\lledd\approx10^{-3}$ -- which is lower than the {\it pre-flare} state of \mysobjagn\ \cite[see][]{ElitzurNetzer2016_true_Sy2}.

The initial cause of the dramatic changes seen here remains to be determined. 
As noted above, the observed timescales are inconsistent with large-scale changes to an AGN-like accretion flow (i.e., a global, sudden change in $\dot{M}$), or to line-of-sight obscuration (i.e., $N_{\rm H}$).

Instead, 
the optical rise time, the peak luminosity, and the following decline are reminiscent of tidal disruption events (TDEs), as illustrated by the comparison to the TDE PS1-10jh \citep{Gezari2012_TDE} and to a generic $t^{-5/3}$ power law \cite[e.g.,][]{Rees1988_TDEs,Phinney1989_TDEs} shown in the bottom panel of \autoref{fig:Swift_LC}.
The mass of the SMBH powering \mysobjagn, of $\mbh \approx 2{\times}10^{7}\,\Msol$, is consistent with what is observed for SMBHs around which TDEs have been seen, but on the high end of the observed distribution \cite[e.g.,][]{Wevers2017_TDE_MBH,VanVelzen2018_TDE_BHMF,Wevers2017_TDE_MBH}.
Associating the changes in \mysobjagn\ with a TDE would be in line with the study of \cite{Merloni2015_CLAGN_TDE}, which argued that another CL-AGN, where the blue continuum and broad-line emission have {\it disappeared} within less than a decade \citep{LaMassa2015_changing}, was driven by a (fading) TDE. 

We note, however, that the freedom in setting the disruption time ($t_{\rm d}$) -- here set to the ATLAS (pre-discovery) detection of the optical transient on 2017 December 23 -- allows to fit such power-law behavior to a vast range of observed light curves. 
Moreover, the lack of extremely broad and strong \HeIIop\ line emission ($\fwhm[\heii] > 10000\,\kms$; $F[\heii] \gg F[\hbeta]$), which is observed in TDEs during the first few months after their detection, as well as the spectral evolution of the continuum and Balmer lines, and the peculiar X-ray light curve, are all different from those seen in tidal disruptions occurring in inactive galaxies \cite[see, e.g.,][]{Gezari2012_TDE,Arcavi2014_TDEs_He,Holoien2014_TDE_AS14ae,Komossa2015_TDE_review,Brown2017_AS14li_longterm,Hung2017_TDE_iPTF16axa}.
While the nuclear transient PS16dtm, a recently claimed TDE in a narrow-line Seyfert 1 AGN \citep{Blanchard2017_PS16dtm}, showed a drop in X-rays following the UV/optical transient that could be similar to what is seen in \mysobjagn, there is no evidence that the X-ray emission has recovered as it did in our case.

A TDE in an AGN could indeed look very different from a simple combination of AGN- and TDE-like observables, given the interaction between the TDE and the pre-existing accretion disk (see the recent study by \citealt{Chan2019_TDEs_in_AGN}).
Detailed modeling of such events is required for comparison with our observations.

Real-time identification and follow-up observations of events such as \mysobjat/\mysobjas\ offer a spectacular opportunity to resolve the mystery of CL-AGN, and to improve our understanding of accretion and BLR physics. 
Fulfilling this potential would require high-cadence optical spectroscopy and complementary multiwavelength data, for a proper sample of such events. 
With the increasing number, cadence, and sky coverage of imaging time-domain surveys (e.g., ASAS-SN, ATLAS, ZTF, LSST) and responsive spectroscopic follow-up programs (e.g., using the Las Cumbres observatory network, ePESSTO, SDSS-V), such events will increasingly be caught in the act.

\acknowledgments
We are grateful to the anonymous referee for helping us improve this manuscript.
We thank Gwen Eadie, Bryce Bolin, and Dino Bekte{\u s}evi{\`c} at UW for assistance with obtaining the APO spectra.
We also thank Hagai Netzer, Dalya Baron, and Robert Antonucci for their insightful and constructive comments. 
R.L.  was supported by the National Key R\&D Program of China (2016YFA0400702) and the National Science Foundation of China (11721303).
I.A. acknowledges support from the Israel Science Foundation (grant number 2108/18).
Support for J.L.P. is provided in part by FONDECYT through the grant 1191038 and by the Ministry of Economy, Development, and Tourism’s Millennium Science Initiative through grant IC120009, awarded to The Millennium Institute of Astrophysics, MAS.
Support for A.V.F.'s research group has been provided by the TABASGO Foundation, the Christopher R. Redlich Fund, and the Miller Institute for Basic Research in Science (U.C. Berkeley). 

This work made use of the MATLAB package for astronomy and astrophysics \citep{Ofek2014_matlab} and Astropy,\footnote{http://www.astropy.org} a community-developed core Python package for Astronomy \citep{Astropy2013,Astropy2018}.
This research also made use of the NASA/IPAC Extragalactic Database (NED), which is operated by the Jet Propulsion Laboratory, California Institute of Technology, under contract with the National Aeronautics and Space Administration.

This work made use of data from 
Las Cumbres Observatory; 
the All Sky Automated Survey for Supernovae (ASAS-SN); 
the Asteroid Terrestrial-impact Last Alert System (ATLAS) project;
the Liverpool Telescope;
the Lick Observatory;
and the W.~M. Keck Observatory.

ATLAS is primarily funded to search for near earth asteroids (NEOs) through NASA grants NN12AR55G, 80NSSC18K0284, and 80NSSC18K1575; byproducts of the NEO search include images and catalogs from the survey area.  
The ATLAS science products have been made possible through the contributions of the University of Hawaii Institute for Astronomy, the Queen's University Belfast, the Space Telescope Science Institute, and the South African Astronomical Observatory.
ASAS-SN is supported by the Gordon and Betty Moore Foundation through grant GBMF5490 to the Ohio State University and NSF grant AST-1515927. %
Development of ASAS-SN has been supported by NSF grant AST-0908816, the Mt.\ Cuba Astronomical Foundation, the Center for Cosmology and AstroParticle Physics at the Ohio State University (CCAPP), the Chinese Academy of Sciences South America Center for Astronomy (CASSACA), the Villum Foundation, and George Skestos. 
We thank the Las Cumbres Observatory and its staff for its continuing support of the ASAS-SN project.  
The Liverpool Telescope is operated on the island of La Palma by Liverpool John Moores University in the Spanish Observatorio del Roque de los Muchachos of the Instituto de Astrofisica de Canarias, with financial support from the UK Science and Technology Facilities Council. 
Research at Lick Observatory 
is partially supported by a generous gift from Google. 
The W. M. Keck Observatory is operated as a scientific partnership among the California Institute of Technology, the University of California, and NASA; the observatory was made possible by the generous financial
support of the W. M. Keck Foundation. 

We finally thank NASA HEASARC for making the {\it Swift} and {\it NICER} data available, and ESA for providing the \xmm\ data.

%

\vspace{5mm}
\facilities{
HST(STIS), 
HST(COS), 
Swift(XRT and UVOT), 
NICER, 
XMM,
ATLAS, 
ASAS-SN, 
MDM Observatory (OSMOS),
Liverpool Telescope(SPRAT),
Lick Observatory (Shane/Kast),
Las Cumbres Observatory (FLOYDS),
Keck (LRIS)
}


\software{astropy \citep{Astropy2013,Astropy2018},  
          SExtractor \citep{BertinArnouts_SE_1996}
          MAAT \citep{Ofek2014_matlab}
}








\end{document}